\documentclass[journal]{IEEEtran}

\usepackage{amsmath, amssymb,graphicx}
\usepackage[scanall]{psfrag}
\newtheorem{theorem}{\bf Theorem}
\newtheorem{proposition}{\bf Proposition}
\newtheorem{lemma}{\bf Lemma}
\newtheorem{corollary}{\bf Corollary}

\begin{document}
\title{Sharp Bounds on the Entropy of the Poisson Law and Related Quantities}

\author{Jos\'{e} A. Adell, Alberto Lekuona, and Yaming~Yu, {\it Member, IEEE}
\thanks{Jos\'{e} A. Adell is with the Departamento de M\'{e}todos Estad\'{i}sticos, Facultad de Ciencias,
Universidad de Zaragoza, Pedro Cerbuna 12, 50009 Zaragoza, Spain
(email: adell@unizar.es).

Alberto Lekuona is with the Departamento de M\'{e}todos
Estad\'{i}sticos, Facultad de Ciencias, Universidad de Zaragoza,
Pedro Cerbuna 12, 50009 Zaragoza, Spain (email:
lekuona@unizar.es).

Yaming Yu is with the Department of Statistics, University of
California, Irvine, CA, 92697-1250, USA (e-mail: yamingy@uci.edu).
}}

\maketitle

\begin{abstract}
One of the difficulties in calculating the capacity of certain
Poisson channels is that $H(\lambda)$, the entropy of the Poisson
distribution with mean $\lambda$, is not available in a simple
form.  In this work we derive upper and lower bounds for
$H(\lambda)$ that are asymptotically tight and easy to compute.
The derivation of such bounds involves only simple probabilistic
and analytic tools.  This complements the asymptotic expansions of
Knessl (1998), Jacquet and Szpankowski (1999), and Flajolet
(1999).  The same method yields tight bounds on the relative
entropy $D(n, p)$ between a binomial and a Poisson, thus refining
the work of Harremo\"{e}s and Ruzankin (2004).  Bounds on the entropy
of the binomial also follow easily.
\end{abstract}



\section{Introduction}\label{sect1}
Unlike the differential entropy for the Gaussian distribution, the
Shannon entropies for many basic discrete distributions, such
as the Poisson, the binomial, or the negative binomial, are ``not
in closed form.''  In the Poisson case, the lack of a simple
analytic expression is seen (\cite{M, LM}) as one of the
obstacles to obtaining the capacity of certain Poisson channels
(\cite{BV, Fr, S, SL}).  Computation of the entropy is also a basic
problem partly motivated by the maximum entropy characterizations
(\cite{SO, M78, H01, T02, J07, Yu08}) of these distributions.

One strategy to make the entropy functions more tractable is to express 
them as integrals (\cite{EBBJ, Kn, M}).  See \cite{GSV, KHJ, MJK} for related integral
representations for entropy-like quantities in the context of Poisson channels.
For the entropy functions themselves, integral representations have been used to
derive asymptotic expansions (\cite{EBBJ, Kn}).  Alternatively, asymptotic
expansions can be obtained using analytic depoissonisation (\cite{JS1, JS2}),
singularity analysis (\cite{Fl}), or local limit theorems (\cite{DV}).

It is obviously desirable to have bounds that accompany asymptotic
expansions, for both theoretical analysis and numerical
computation.  This is especially true for quantities such as the
entropy of the Poisson law, which may be used, e.g., in capacity
calculations for discrete-time Poisson channels (\cite{M},
\cite{LM}).  Part of this work aims to derive tight bounds on the
entropy for fundamental distributions such as the Poisson and the
binomial.  Our results are expressed in terms of two sequences of
lower and upper bounds, and have the following features.
\begin{itemize}
\item The sequence of upper bounds and the sequence of lower
bounds each gives a full asymptotic expansion for the entropy.  In
other words the bounds are asymptotically tight. \item The bounds
are derived from familiar quantities such as the central moments
of the Poisson, and are in a form simple enough for both
theoretical analysis and numerical computation. \item The
derivation, which involves only real analysis, is elementary.
(Note that the asymptotic expansions of Knessl \cite{Kn} are also real-analytic.)
\end{itemize}

Denote by $\mathbf{Z}_+=\{0, 1, 2, \ldots\}$ and by
$\mathbf{N}=\mathbf{Z}_+\setminus \{0\}$. As usual, the Shannon
entropy for a discrete random variable $X$ on $\mathbf{Z}_+$ with
mass function $f_i=\Pr(X=i)$ is defined as
$$H(X)=H(f)=\sum_{i=0}^\infty -f_i\log f_i,$$
where we use the natural logarithm and obey the convention $0\log
0=0$. Throughout we use $N_\lambda$ to denote a Poisson random
variable with mean $\lambda$, i.e., the mass function is
$$\Pr(N_\lambda=j)=\frac{e^{-\lambda}\lambda^j}{j!},\quad j\in
\mathbf{Z}_+.$$ 
We write $H(\lambda)=H(N_\lambda)$ for
simplicity.  The best known bound on $H(\lambda)$ is perhaps
\begin{equation}
\label{ctbd} H(\lambda)\leq \frac{1}{2}\log\left(2\pi
e\left(\lambda+\frac{1}{12}\right)\right),
\end{equation}
which is obtained by bounding the differential entropy of
$N_\lambda+U$ where $U$ is an independent random variable
uniformly distributed on $(0, 1)$ (\cite{CT}, Theorem 8.6.5).
While (\ref{ctbd}) is simple in form and reasonably accurate, it
lacks a corresponding lower bound, and does not extend easily to
capture higher order terms in the expansion of $H(\lambda)$.  As a
remedy we shall derive, for each $m\geq 1$, a double inequality of
the form
\begin{align}\label{main}
\nonumber \sum_{k=m}^{2m} \frac{A(m, k)}{\lambda^k} &\leq
H(\lambda)-\frac{1}{2}\log (2\pi \lambda)
-\frac{1}{2}-\sum_{k=1}^{m-1} \frac{b(m,k)}{\lambda^k}\\
& \leq \sum_{k=m}^{2m-1} \frac{B(m, k)}{\lambda^k},
\end{align}
where $A(m, k)$, $b(m, k)$, and $B(m,k)$ are explicit constants
($\sum_1^0\equiv 0$). In other words, for each $m\geq 1$, we give
a finite asymptotic expansion in powers of $\lambda^{-1}$ with $m$
exact terms and explicit lower and upper bounds of the order of
$\lambda ^{-m}$.

In Section \ref{sect2} we derive (in an equivalent form) the
double inequality (\ref{main}). The key steps are
\begin{itemize}
\item an integral representation that relates $H(\lambda)$ to the
simpler quantity $E[\log(N_\lambda+1)]$;
\item bounds on
$E[\log(N_\lambda+1)]$ in terms of polynomials in $\lambda^{-1}$,
which translate easily to bounds on $H(\lambda)$.
\end{itemize}
Note that (\ref{main}) is only effective for large $\lambda$. We
also obtain bounds on $H(\lambda)$ in terms of polynomials in
$\lambda$, which work for small $\lambda$.

Besides $H(\lambda)$, we also consider bounds on the relative
entropy between a binomial and a Poisson, thus obtaining a version
of ``the law of small numbers'' that refines the results of
Harremo\"{e}s and Ruzankin \cite{HR}.  While these are of theoretical
interest, they also lead to new bounds for the entropy of the
binomial.  As usual, for two random variables $X$ and $Y$ on
$\mathbf{Z}_+$ with mass functions $f$ and $g$ respectively, the
relative entropy is defined as
$$D(X\|Y)=D(f\|g)=\sum_{j=0}^\infty f_j\log \frac{f_j}{g_j}.$$
By convention $0\log(0/0)=0,$ and $D(f\|g)=\infty$ if $f$ assigns
mass outside of the support of $g$.  Throughout we let $B_{n,p}$
be a binomial random variable with mass function
$$\Pr(B_{n,p}=k)=\binom{n}{k} p^k q^{n-k},\quad k=0, 1,\ldots, n,$$
where $q\equiv 1-p,\ p\in (0,1)$ and $n \in \mathbf{N}$.  We
consider
$$D(n, p)=D(B_{n,p}\|N_{np}),$$
i.e., the relative entropy between $B_{n,p}$ and $N_{np}$, and
derive bounds on $D(n, p)$ using similar techniques.  Bounds on
the entropy of the binomial $H(B_{n,p})$ are obtained as a
corollary.  

Sections III and IV contain proofs of the main results.  We conclude 
with a short discussion on possible extensions in Section~\ref{sect4}.

\section{Main Results}\label{sect2}
\subsection{Sharp Bounds on $H(\lambda)$}
We first present a class of double inequalities for $H(\lambda)$
that is effective for small $\lambda$ (say $\lambda\leq 1$), but
valid for all $\lambda\geq 0$.

\begin{theorem}
\label{thm1} For any $\lambda\geq 0$ and $m=1, 2, \ldots$, we have
$$\sum_{k=2}^{2m+1} \frac{c(k)}{k!} \lambda^k \leq
H(\lambda)+\lambda \log \lambda -\lambda \leq \sum_{k=2}^{2m}
\frac{c(k)}{k!} \lambda^k,$$ where
$$c(k)=\sum_{j=0}^{k-1} (-1)^{k-1-j} \binom{k-1}{j} \log(j+1),\quad k=2, 3, \ldots.$$
\end{theorem}

For fixed $m$, the two bounds given by Theorem~\ref{thm1} 
differ by $O(\lambda^{2m+1})$.  Hence they are most effective 
when $\lambda$ is small.  Moreover, inspection of the proof 
(Section~III, Part B) shows that, for $0\leq \lambda\leq 1$,
both the upper and lower bounds in Theorem~\ref{thm1} converge to
$H(\lambda) + \lambda \log \lambda -\lambda$ as $m\to\infty$. 

In what follows, the $k$th central moment of the Poisson
distribution
$$\mu_k(s)\equiv E[(N_s-s)^k]$$
plays an important role.  The first few values of $\mu_k(s)$ are
$\mu_0(s)=1,\ \mu_1(s)=0$, and
$$\mu_2(s)=\mu_3(s)=s,\ \mu_4(s)=3s^2+s,\ \mu_5(s)=10s^2+s.$$
They obey the well-known recursion (\cite{JKK}, p. 162)
\begin{equation}
\label{rec}
\mu_k(s)=s\sum_{j=0}^{k-2} \binom{k-1}{j}
\mu_j(s),\quad k\geq 2,
\end{equation}
from which it is easy to show that, for $k\geq 2$, $\mu_{k}(s)$ is
a polynomial in $s$ of degree $\lfloor k/2 \rfloor$, where
$\lfloor x\rfloor$ denotes the integer part of $x$.

In contrast to Theorem \ref{thm1}, Theorem \ref{thm2} is most
effective for large $\lambda$.
\begin{theorem}
\label{thm2} For any $\lambda> 0$ and $m=1, 2,\ldots,$ we have
$$
-r_m(\lambda )\leq H(\lambda)-\frac{1}{2}\log (2\pi \lambda)
-\frac{1}{2}-\beta_m (\lambda )\leq 0,
$$
where
\begin{equation}
\label{beta} \beta_m(\lambda )=\int_\lambda^\infty
\left(\sum_{j=3}^{2m+1} \frac{(-1)^{j-1} \mu_j(s)}{j(j-1)
s^j}\right)\, {\rm d}s= \sum_{k=1}^{2m-1}\frac{b(m,k)}{\lambda ^k}
\end{equation}
and
\begin{equation}
\label{br} r_m(\lambda )=\int_\lambda ^\infty
\frac{\mu_{2m+2}(s)}{(2m+1)s^{2m+2}}\, {\rm d}s=
\sum_{k=m}^{2m}\frac{a(m,k)}{\lambda ^k}.
\end{equation}
\end{theorem}

Let us note that the seemingly
cumbersome expressions (\ref{beta}) and (\ref{br}) are actually
quite easy to handle.  For $j\geq 3$, the $j$th central moment
$\mu_j(s)$ is a polynomial in $s$ of degree $\lfloor j/2 \rfloor$.
This, together with (\ref{rec}), shows that the integrand in
(\ref{beta}) is a polynomial in $s^{-1}$, with powers going from
$s^{-2}$ to $s^{-2m}$. Similar statements hold for the integrand
in (\ref{br}).  Hence the constants $a(m,k)$ and $b(m,k)$ are
obtained after straightforward integration; see Table I for their
values for small $m$.  In particular, for $m=2$ we have
\begin{align*}
-\frac{31}{24
\lambda^2}-\frac{33}{20\lambda^3}-\frac{1}{20\lambda^4}
\leq H(\lambda) &-\frac{1}{2}\log(2\pi\lambda)-\frac{1}{2}+\frac{1}{12\lambda}\\
&\leq \frac{5}{24\lambda^2}+\frac{1}{60\lambda^3}.
\end{align*}

We emphasize that the constants $b(m,k)$, $1\leq k\leq m-1$, are
exact in the full asymptotic expansion of $H(\lambda )$, since (\ref{br}) gives 
$r_m(\lambda)=O(\lambda^{-m})$.  For
example, we have $b(3,1)=-1/12$ and $b(3,2)=-1/24$ from Table I,
and hence
$$H(\lambda)=\frac{1}{2}\log (2\pi \lambda) +\frac{1}{2}
-\frac{1}{12\lambda}-\frac{1}{24\lambda^2}+O(\lambda^{-3}),
$$
which agrees with the leading terms given by, e.g., Knessl
(\cite{Kn}, Theorem 2).

\begin{table} \begin{center}
\caption{Values of $a(m, k)$ and $b(m, k)$ for $m=1,2,3,4$.}
\begin{tabular}{l|rrrr}
\ & \multicolumn{4}{c}{$a(m, k)$}   \\
\hline k &   $m=1$            & $m=2$  & $m=3$  & $m=4$  \\
\hline $1$ & $1$   &  \   &   \  &  \\
       $2$ & $1/6$  & $3/2$    &   \  &  \\
       $3$ & \        & $5/3$   &   $5$ & \\
       $4$ & \        & $1/20$    &   $35/2$  &$105/4$\\
       $5$ & \        & \          &   $17/5$ & $210$\\
       $6$ & \        & \          &   $1/42$   & $2275/18$ \\
       $7$ & \        & \          &            & $167/21$\\
       $8$ & \        & \          &            & $1/72$\\
       \hline
  $k$ &  \multicolumn{4}{c}{$b(m, k)$} \\
\hline $1$ & $1/6$    & $-1/12$  &  $-1/12$  &  $-1/12$\\
       $2$ & \        & $5/24$   &  $-1/24$   & $-1/24$\\
       $3$ & \        & $1/60$   &  $103/180$ & $-19/360$\\
       $4$ & \        & \        &  $13/40$   & $201/80$\\
       $5$ & \        & \        &  $1/210$  & $12367/2520$\\
       $6$ & \        & \        & \         & $571/1008$\\
       $7$ & \        & \        & \         & $1/504$\\
\hline
\end{tabular}
\end{center}
\end{table}

Bounds on $H(\lambda)$ given by Theorem \ref{thm2} are illustrated
in Fig.~1.  As $\lambda$ increases from $10$ to $20$, the gap
between upper and lower bounds, $r_m(\lambda)$, decreases
\begin{itemize}
\item from $\approx 0.1$ to $\approx 0.05$ with $m=1$, \item from
$\approx 0.017$ to $\approx 0.004$ with $m=2$, and \item from
$\approx 0.0068$ to $\approx 0.00074$ with $m=3$.
\end{itemize}
In short, for moderate $\lambda$, the bounds are already quite
accurate with $m$ as small as $3$, and, as expected, the accuracy
improves as $\lambda$ increases.

\begin{figure}
\begin{center}
\psfrag{bounds on H(lambda)}{\small bounds on $H(\lambda)$}
\psfrag{gaps between the two bounds}{\small gaps between bounds} 
\psfrag{lam}{\small $\lambda$}
\includegraphics[width=2.3in, height=3.4in, angle=270]{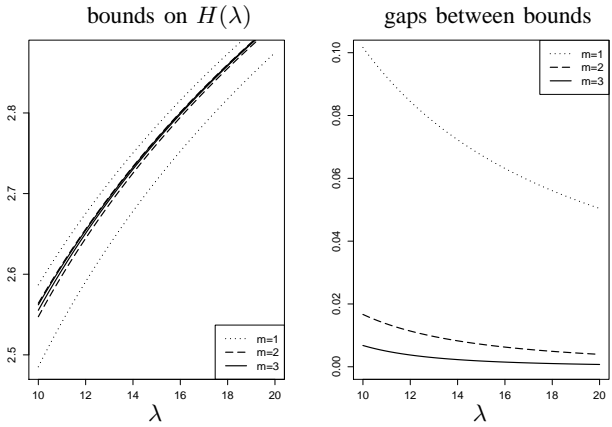}
\end{center}
\caption{Bounds on $H(\lambda)$ (left) and differences between
upper and lower bounds (right) given by Theorem \ref{thm2} with
$m=1, 2, 3$.}
\end{figure}

\subsection{Exact Formulae and Sharp Bounds for $D(n,p)$}

While bounds on Poisson convergence are often stated in terms of
the total variation distance (\cite{BHJ, AA}), those based on the
relative entropy can also be quite effective (\cite{KHJ, HR}).  As
a discrepancy measure, relative entropy occurs naturally in
contexts such as hypothesis testing.  In this section we consider
$D(n,p)\equiv D(B_{n,p}\|N_{np})$, and study its higher-order asymptotic
behavior.  Our main result, Theorem \ref{thm4}, may
be regarded as a refinement of that of Harremo\"{e}s and Ruzankin
\cite{HR}.  As a by-product, we also obtain bounds on the entropy
of the binomial that parallel Theorem \ref{thm2}.

Analogous to Theorem \ref{thm1}, we have the following exact
expansion for $D(n,p)$ as a function of $p$.

\begin{theorem}
\label{thm3} Fix $n\in \mathbf{N}$. For $p\in [0,1]$ we have
$$D(n,p)=n(p+q\log q)+\sum_{k=2}^n \binom{n}{k} \tilde{c}(k) p^k,$$ where
$$\tilde{c}(k)=\sum_{j=0}^{k-1}(-1)^{k-1-j}\binom{k-1}{j}\log (n-j),\ k=2,\ldots ,n.$$
\end{theorem}

In the following result, the $k$th central moment of $B_{n,p}$,
$$\mu_k(n, p)\equiv E[(B_{n,p}-np)^k],$$
plays an important role.  The first few values of $\mu_k(n, p)$
are
$$\mu_0(n,p)=1,\ \mu_1(n, p)=0,\ \mu_2(n, p)=pqn,$$
$$\mu_3(n,p)=(q-p)pqn,\ \mu_4(n,p)=3(pqn)^2+(1-6pq)pqn.$$
We write $q\equiv 1-p$ throughout.

Analogous to Theorem \ref{thm2}, Theorem \ref{thm4} gives bounds
on $D(n,p)$ that are effective for large $n$.
\begin{theorem}
\label{thm4} For $n,\, m\in \mathbf{N}$ and $p\in (0,1)$, we have
$$0\leq D(n, p) +\frac{p+\log q}{2}-\tilde{\beta}_m(n,p)\leq \tilde{r}_m(n,p),$$
where
\begin{equation*}
\tilde{\beta}_m(n,p)=n\int_q^1 \sum_{j=3}^{2m+1} \frac{(-1)^j
\mu_j(n, s)}{j(j-1)(ns)^j}\, {\rm d}s=\sum_{k=1}^{2m-1}
\frac{\tilde{b}(m,k;p)}{n^k}
\end{equation*}
and
\begin{equation*}
\tilde{r}_m(n,p)=n\int_q^1 \frac{ \mu_{2m+2}(n,
s)}{(2m+1)(ns)^{2m+2}}\, {\rm d} s
=\sum_{k=m}^{2m}\frac{\tilde{a}(m,k;p)}{n^k}.
\end{equation*}
\end{theorem}

The integrals that define $\tilde{\beta}_m(n, p)$ and $\tilde{r}_m(n, p)$ 
are easy to calculate, because $\mu_j(n, s)$ is a polynomial in $s$.  We 
obtain the coefficients $\tilde{b}(m, k; p)$ and $\tilde{a}(m, k; p)$ after 
integrating and assembling the results in powers of $n^{-1}$.  For example, 
we have ($m=1$)
\begin{align*}
\tilde{b}(1, 1; p) &=-\frac{\log q}{2}
+\frac{q}{3}-\frac{1}{6q}-\frac{1}{6},\\
\tilde{a}(1, 1; p) &=2\log q-q+\frac{1}{q},\\
\tilde{a}(1, 2; p) &=-4\log
q+2q-\frac{7}{3q}+\frac{1}{6q^2}+\frac{1}{6},
\end{align*}
and ($m=2$)
\begin{align}
\label{tildeb21} \tilde{b}(2, 1; p) &=\frac{p^2}{12q},\\
\nonumber \tilde{b}(2, 2; p) &=\frac{3}{2}\log q-\frac{1}{2}\,
q+\frac{17}{12q}
-\frac{5}{24q^2}-\frac{17}{24},\\
\nonumber \tilde{b}(2, 3; p) &=-3\log q+\frac{6}{5}\,
q-\frac{5}{2q} +\frac{3}{8q^2}-\frac{1}{60q^3}+\frac{113}{120},\\
\nonumber \tilde{a}(2, 2; p) &=-9\log q +3\, q-\frac{9}{q}
+\frac{3}{2q^2}+\frac{9}{2},\\
\nonumber \tilde{a}(2, 3; p) &=78\log q -26\, q+\frac{83}{q}
-\frac{18}{q^2}+\frac{5}{3q^3}-\frac{122}{3},\\
\nonumber \tilde{a}(2, 4; p) &=-72\log q+24\, q-\frac{78}{q}
+\frac{18}{q^2}-\frac{31}{15q^3}\\
\nonumber &\quad +\frac{1}{20q^4}+\frac{2281}{60}.
\end{align}

As in Theorem \ref{thm2}, we emphasize that, for each $m\geq 2$,
the constants $\tilde{b}(m,k;p)$, $1\leq k\leq m-1$, are exact in
the full asymptotic expansion of $D(n,p)$ (for fixed $p$ as $n\to
\infty$).  In particular, from (\ref{tildeb21}) we get
$$D(n,p)=-\frac{p+\log q}{2}+ \frac{p^2}{12qn} +O(n^{-2}),$$
for fixed $p\in (0,1)$.  We can also estimate the rate at which
$D(n, \lambda/n)$ decreases to zero, for fixed $\lambda>0$ as
$n\to\infty$, which corresponds to the usual binomial-to-Poisson
convergence.  Indeed, setting $m=1$ and $p=\lambda/n$, Theorem
\ref{thm4} yields, after routine calculations,
$$D(n, \lambda/n)=\frac{\lambda^2}{4n^2}+O(n^{-3}),$$
which can be further refined by using larger $m$.

Such results are related to those of Harremo\"{e}s and Ruzankin
\cite{HR}, who give several bounds on $D(n,p)$ after detailed
analyses on inequalities involving Stirling numbers.  Theorem~\ref{thm4} 
may be regarded as a refinement in that, by letting $m$
be arbitrary, it can give a full asymptotic expansion of $D(n,
p)$, with computable bounds.  Our derivation is also simpler (see
Section~IV).

Let us denote the entropy of the binomial by $H(n, p)=H(B_{n,p})$.
In the special case $p=1/2$, $H(n, p)$ appears as the sum capacity
of a noiseless $n$-user binary adder channel as analyzed by Chang
and Weldon \cite{CW}, who also provide simple bounds on $H(n, p)$.
For general $p$, asymptotic expansions for $H(n, p)$ have been
obtained by Jacquet and Szpankowski \cite{JS2} and Knessl
\cite{Kn} (see also Flajolet \cite{Fl}).  We present sharp bounds
on $H(n,p)$ that complement such expansions.  As it turns out, due
to an elementary identity (see \eqref{eq4} below), bounds on $D(n,p)$
obtained in Theorem \ref{thm4} translate directly to those on
$H(n, p)$, thus simplifying our analysis.

\begin{corollary}
\label{coro} Let $n, m\in \mathbf{N}$ and $p\in (0,1)$.  Define
$$\tilde{H}(n, p)=H(n, p)-\log n!+n\log n-n -\frac{1+\log (pq)}{2}.$$
Then
\begin{align*}
0 &\leq -\tilde{H}(n, p)-\sum_{k=1}^{2m-1} \frac{\tilde{b}(m, k;
p)
+\tilde{b}(m, k; q)}{n^k} \\
&\leq \sum_{k=m}^{2m} \frac{\tilde{a}(m,k; p)+\tilde{a}(m, k;
q)}{n^k}\ ,
\end{align*}
where $\tilde{a}(m, k; p)$ and $\tilde{b}(m, k; p)$ are defined as
in Theorem \ref{thm4}.
\end{corollary}

Bounds on $H(n,p)$ can also be expressed in terms of $\log n$ and
$n^{-k},\ k=1, 2,\ldots,$ via familiar bounds on $\log n!$.  For
example, taking $m=1$ in Corollary \ref{coro}, and using (see
\cite{AS}, 6.1.42)
$$\frac{1}{12n}-\frac{1}{360 n^3}<\log n!-n\log n +n
-\frac{1}{2}\log(2\pi n)<\frac{1}{12n},$$ we get
\begin{equation}
\label{bibd} \frac{C_1}{n}+\frac{C_2}{n^2}+\frac{C_3}{n^3}<H(n,p)-
\frac{1}{2}\log(2\pi npq)-\frac{1}{2} <\frac{C_4}{n},
\end{equation}
where
\begin{align*}
C_1 &=\frac{13}{12}-\frac{3\log(pq)}{2}-\frac{5}{6pq},\\
C_2 &=-\frac{7}{3}+4\log(pq)+\frac{8}{3pq} -\frac{1}{6(pq)^2},\\
C_3 &=-\frac{1}{360},\\
C_4 &=\frac{1}{12}+\frac{\log(pq)}{2}+\frac{1}{6pq}.
\end{align*}
To relate (\ref{bibd}) to the Poisson case, i.e., Theorem
\ref{thm2}, let us fix $\lambda>0$ and set $p=\lambda/n$.  Then,
as $n\to\infty$, we have $H(n,p)\to H(\lambda)$, and in the limit
(\ref{bibd}) becomes
$$-\frac{5}{6 \lambda}-\frac{1}{6\lambda^2} \leq H(\lambda)
-\frac{1}{2}\log(2\pi\lambda)-\frac{1}{2} \leq
\frac{1}{6\lambda},$$ which is precisely Theorem \ref{thm2} with
$m=1$.  In general ($m\geq 1$), Theorem \ref{thm2} may be seen as
a limiting case of Corollary \ref{coro}.

As before, for each $m\geq 2$, the coefficients $\tilde{b}(m, k;
p)+\tilde{b}(m, k; q)$, $k=1, \ldots, m-1$, are exact in the
asymptotic expansion of $\tilde{H}(n,p)$ (for fixed $p$ as $n\to
\infty$).  For large $n$, we may choose larger $m$ to improve the
accuracy.

\section{Proofs of Theorems \ref{thm1} and \ref{thm2}}\label{sect3}

\subsection{Preliminaries}

Given a function $\phi:\, \mathbf{R}\to \mathbf{R}$, the $m$th
forward difference of $\phi$ is defined recursively by
$$\Delta^0\phi (x) =\phi (x),\ \ \Delta^{m+1}
\phi (x)=\Delta^m \phi (x+1)-\Delta^m \phi (x),$$ for any $x\in
\mathbf{R}$ and $m\in \mathbf{Z}_+$. Equivalently, we have 
\begin{equation}
\label{delta} \Delta^m \phi(x)=\sum_{j=0}^m (-1)^{m-j}
\binom{m}{j} \phi(x+j),\ x\in \mathbf{R},\ m\in \mathbf{Z}_+.
\end{equation}
(This definition extends to functions defined only on
$\mathbf{Z}_+$.) As usual, $\phi^{(m)}$ stands for the $m$th
derivative of $\phi$. If $\phi$ is infinitely differentiable, then
\begin{equation}
\label{poidel} \Delta^m \phi(x)=E [\phi^{(m)}(x+S_m)],
\end{equation}
where $S_m=U_1+\cdots+U_m$ and $(U_k)_{k\geq 1}$ is a sequence of
independent and identically distributed (i.i.d.) random variables
having the uniform distribution on $[0,1]$ (cf. \cite{AL}, Eqn.\
(2.7)).

On the other hand, two special properties of the Poisson
distribution are
\begin{equation}
\label{poi} \lambda E[\phi (N_\lambda +1)]=E[N_\lambda \phi
(N_\lambda )],\quad \lambda \geq 0,
\end{equation}
and (see \cite{AL}, Theorem 2.1, for example)
\begin{equation}
\label{deri} \frac{{\rm d}^m E[\phi(N_\lambda)]}{{\rm d} \lambda^m}
=E[\Delta^m \phi(N_\lambda)], \quad \lambda\geq 0,\ m\in
\mathbf{Z}_+.
\end{equation}
In both (\ref{poi}) and (\ref{deri}), $\phi:\, \mathbf{Z}_+\to
\mathbf{R}$ is an arbitrary function for which the relevant
expectations exist. From (\ref{deri}) we have the Taylor formula
\begin{align}
\nonumber
E[\phi(N_\lambda)]=&\sum_{k=0}^m \frac{\Delta^k \phi(0)}{k!} \lambda^k \\
\label{taylor1} &+\frac{1}{m!} \int_0^\lambda (\lambda-u)^m
E[\Delta^{m+1} \phi(N_u)]\, {\rm d}u,
\end{align}
for any $\lambda\geq 0$ and $m\in \mathbf{Z}_+$.

\subsection{Proof of Theorem \ref{thm1}}

Let $\lambda\geq 0$.  We have
\begin{equation}
\label{eq5}
H(\lambda)=\lambda-\lambda\log \lambda +E[\log
N_\lambda!]
\end{equation}
by definition.  Applying (\ref{taylor1}) to the function
$\phi(k)=\log k!,\ k\in \mathbf{Z}_+$, we obtain
\begin{align}
\nonumber H(\lambda)+\lambda\log \lambda -\lambda =&\sum_{k=2}^m
\frac{\Delta^k
\phi(0)}{k!} \lambda^k\\
\label{taylorh}
 &+\frac{1}{m!}\int_0^\lambda (\lambda-u)^m E[\Delta^{m+1} \phi(N_u)]\, {\rm d}u,
 \end{align}
for any $m=2, 3, \ldots,$ since $\Delta^0 \phi(0)=\Delta^1
\phi(0)=0$.

On the other hand, letting $g(x)=\log(x+1),\ x\geq 0$, and using
(\ref{delta}) and (\ref{poidel}), we can write
\begin{align}
\nonumber
\Delta^{m+1} \phi(i)= &\Delta^m g(i)\\
\label{alternate}
=&(-1)^{m-1} E \left [\frac{(m-1)!}{(i+1+S_m)^m}\right ]\\
\nonumber =& \sum_{j=0}^m (-1)^{m-j} \binom{m}{j} \log(i+1+j),
\end{align}
where $i\in \mathbf{Z}_+$ and $m=1, 2, \ldots$. Therefore,
\begin{equation}
\label{eq1} \Delta^k \phi (0)=\sum_{j=0}^{k-1}
(-1)^{k-1-j}\binom{k-1}{j}\log (j+1)=c(k),
\end{equation}
for $k=2,3,\ldots$. The conclusion follows from (\ref{taylorh})
and \eqref{eq1} by noting that, based on (\ref{alternate}), the
integral in (\ref{taylorh}) alternates in sign for $m=1,2,
\ldots$.

\subsection{Proof of Theorem \ref{thm2}}

Some auxiliary results are needed in the proof of Theorem
\ref{thm2}. In Lemmas \ref{lem1} and \ref{lem2} we denote
\begin{equation}
\label{htilde}
\tilde{H}(\lambda)=H(\lambda)-\frac{1}{2}\log(2\pi\lambda)-\frac{1}{2},\quad
\lambda>0,
\end{equation}
for convenience.

\begin{lemma}
\label{lem1} We have $\tilde{H}(\lambda)  =O(\lambda^{-1}),$ as
$\lambda\to\infty$.
\end{lemma}

\begin{IEEEproof}
See \cite{EBBJ} or \cite{Kn}.
\end{IEEEproof}

Lemma \ref{lem2} expresses the quantity of interest in terms of an
easier quantity, $E[\log(N_s+1)]$.
\begin{lemma}
\label{lem2} For $\lambda >0$ we have
\begin{equation*}
\tilde{H}(\lambda)=\int_\lambda^\infty \left(\frac{1}{2s}-E \left [\log
\frac{N_s+1}{s} \right ]\right)\, {\rm d}s.
\end{equation*}
\end{lemma}

\begin{IEEEproof}
Recalling \eqref{eq5} and using (\ref{deri}) with $\phi(k)=\log
k!,\ k\in \mathbf{Z}_+$ and $m=1$, we obtain from (\ref{htilde})
$$\tilde{H}'(\lambda) =E [\log(N_\lambda+1)]-\log \lambda -\frac{1}{2\lambda}.
$$
By Lemma \ref{lem1}, $\tilde{H}(\infty)=0$, and the claim follows.
\end{IEEEproof}

The next result presents sharp bounds on $E[\log(N_s+1)]$.
\begin{proposition}
\label{prop1} For $s>0$ and $m\in \mathbf{N}$, we have
\begin{align*}
0\leq E\left[\log\frac{N_s+1}{s}\right ] -\sum_{k=2}^{2m+1} \frac{(-1)^k
\mu_k(s)}{k(k-1) s^k}\leq \frac{\mu_{2m+2}(s)}{(2m+1) s^{2m+2}}.
\end{align*}
\end{proposition}
\begin{IEEEproof}
By (\ref{poi}) we have
\begin{equation}
\label{eq2} E\left [ \log \frac{N_s+1}{s} \right ]= \frac{1}{s}
\left ( E [N_s\log N_s]-s\log s\right ).
\end{equation}
Taking into account that $E[N_s]=s$, the lower bound follows from
\eqref{eq2} and the inequality (upon letting $x=N_s$)
\begin{equation}
\label{xlogxbd} x\log x-s\log s\geq (\log s+1)(x-s)
+\sum_{k=2}^{2m+1}\frac{(-1)^k (x-s)^k}{k(k-1) s^{k-1}}\ ,
\end{equation}
which holds for $x\geq 0,\ s>0.$

On the other hand, using the inequality
\begin{equation}
\label{logbd} \log(1+x)\leq \sum_{k=1}^{2m+1}\frac{(-1)^{k-1}}{k}
x^k,\quad x>-1,
\end{equation}
with $x=(N_s-s+1)/s$, we obtain
\begin{equation}
\label{logbd2} E\left [\log\frac{N_s+1}{s}\right ]\leq \sum_{k=1}^{2m+1}
\frac{(-1)^{k-1}E[(N_s-s+1)^k]}{ks^k}.
\end{equation}
Again by (\ref{poi}), we have
$$sE[(N_s-s+1)^k]=E[N_s(N_s-s)^k]=\mu_{k+1}(s)+s\mu_k(s).$$
Hence the right-hand side of (\ref{logbd2}) is equal to
$$\frac{\mu_{2m+2}(s)}{(2m+1)s^{2m+2}}+\sum_{k=2}^{2m+1}
\frac{(-1)^k\mu_k(s)}{k(k-1)s^k},
$$
which proves the upper bound.
\end{IEEEproof}

{\bf Remark}. The quantity $E[\log(N_s+1)]$ already appears as
Example 2 of Entry 10 in Chapter 3 of Ramanujan's second notebook
(\cite{B}).  Proposition \ref{prop1} gives a finite expansion of
$E[\log(N_s+1)]$ with an explicit upper bound of the order of
$s^{-(m+1)}$.  See \cite{Yu1} for related work on this particular
entry of Ramanujan.

\begin{IEEEproof}[Proof of Theorem \ref{thm2}]
The claim follows directly from Lemma \ref{lem2} and Proposition
\ref{prop1}, upon noting $\mu_2(s)=s$.
\end{IEEEproof}

\section{Proofs of Theorems \ref{thm3} and \ref{thm4} and Corollary \ref{coro}}

\subsection{Preliminaries}
Let us recall two special properties of the binomial variable
$B_{n,p}$.  For $n\in \mathbf{N}$, $p\in [0,1]$, and any function
$\phi:\ \{0, 1, \ldots, n\}\to \mathbf{R}$, we have
\begin{equation}
\label{binomial} npE[\phi(B_{n-1,p}+1)]=E[B_{n,p}\phi(B_{n,p})],
\end{equation}
as well as (see \cite{AA}, Theorem 1, for example)
\begin{align}
\label{bi_deri}
\frac{{\rm d}^k E[\phi(B_{n,p})]}{{\rm d}p^{k}}  &=k!\binom{n}{k}
E[\Delta^k \phi(B_{n-k, p})], \quad k=0, 1, \ldots, n;\\
\nonumber
\frac{{\rm d}^k E[\phi(B_{n,p})]}{{\rm d}p^{k}} &=0, \quad k=n+1, n+2, \ldots.
\end{align}
From (\ref{bi_deri}) we obtain the Taylor formula
\begin{equation}
\label{taylor_bi}
E[\phi(B_{n,p})]=\sum_{k=0}^n \binom{n}{k} E[\Delta^k\phi(B_{n-k, t})](p-t)^k, 
\end{equation}
where $p, t\in [0,1]$.

\subsection{Proof of Theorem \ref{thm3}}

By direct calculation
\begin{equation}
\label{dnp} D(n,p)=n(p+q\log q)-np\log n +E\left [\log\frac{n!}{(n-B_{n,
p})!}\right ].
\end{equation}
Using (\ref{taylor_bi}) with $\phi(l)=\log(n!/(n-l)!),\ l=0,
\ldots, n$ and $t=0$, we see that the expectation in (\ref{dnp})
is equal to
\begin{equation}
\label{bi_expect} E[\phi(B_{n,p})]=\sum_{k=0}^n \binom{n}{k}
\Delta^k \phi(0) p^k,\quad p\in [0,1].
\end{equation}
Denote by $g(l)=\log(n-l)$, $l=0,1,\ldots ,n-1$. Observe that
$$\Delta^k \phi=\Delta^{k-1} g,\ k\in \mathbf{N},\quad \Delta^0
\phi(0)=0, \quad \Delta^1 \phi(0)=\log n.$$ Therefore, we have
from \eqref{delta}
\begin{align}
\label{eq3} \nonumber\Delta^k \phi (0)=&\Delta^{k-1}g(0)\\
=&\sum_{j=0}^{k-1} (-1)^{k-1-j} \binom{k-1}{j}\log
(n-j)=\tilde{c}(k),
\end{align}
for $k=2,\ldots ,n$. The claim follows from (\ref{dnp}),
(\ref{bi_expect}), and \eqref{eq3}.\hfill$\blacksquare$

\subsection{Proof of Theorem \ref{thm4}}

The following integral representation is crucial in the proof of Theorem \ref{thm4}.
\begin{lemma} \label{lem3}
For any $p\in [0,1]$ we have
$$D(n,p)=n\int_q^1 E\left[ \log\frac{B_{n-1, s}+1}{ns}\right ]\, {\rm d}s.$$
\end{lemma}
\begin{IEEEproof}
Differentiating (\ref{dnp}) once and applying (\ref{bi_deri}) with
$k=1$ and $\phi(l)=\log(n!/(n-l)!)$, we get
$$\frac{{\rm d} D(n, p)}{{\rm d} p}=nE[\log(n-B_{n-1, p})] -n\log(nq).$$
Since $D(n, 0)=0$, we have
$$D(n,p) =n\int_0^p \left (E[\log(n-B_{n-1, t})] -\log(n(1-t))\right)\, {\rm d}t.
$$
Note that $n-B_{n-1, t}$ and $B_{n-1, 1-t}+1$ have the same
distribution. The claim follows by a change of variables $s=1-t$.
\end{IEEEproof}

Proposition \ref{propbi} gives upper and lower bounds on the key
quantity $E[\log((B_{n-1, s}+1)/(ns))]$.  This parallels Proposition~\ref{prop1}.
\begin{proposition}
\label{propbi} For any $m\in \mathbf{N}$ and $s\in (0,1)$ we have
\begin{align*}
0 &\leq E\left[\log\frac{B_{n-1, s}+1}{ns}\right ]-\sum_{k=2}^{2m+1} \frac{(-1)^k
\mu_k(n, s)}{k(k-1)(ns)^k} \\
&\leq \frac{\mu_{2m+2}(n, s)}{(2m+1)(ns)^{2m+2}}.
\end{align*}
\end{proposition}
\begin{IEEEproof}
By (\ref{binomial}) we have
$$nsE[\log(B_{n-1, s}+1)]=E[B_{n,s}\log(B_{n,s})].$$
The lower bound follows from this and (\ref{xlogxbd}).

As in the proof of Lemma \ref{lem2}, using
(\ref{logbd}) with
$$x=\frac{B_{n-1, s}+1-ns}{ns},$$
we obtain
\begin{equation}
\label{bound}
E\left [\log \frac{B_{n-1,s}+1}{ns}\right ]\leq \sum_{k=1}^{2m+1}
\frac{(-1)^{k-1}E[(B_{n-1,s}-ns+1)^k]}{k(ns)^k}.
\end{equation}
Again by (\ref{binomial}), we have
\begin{align*}
nsE[(B_{n-1,s}-ns+1)^k] &=E[B_{n,s}(B_{n,s}-ns)^k]\\
&=\mu_{k+1}(n,s)+ns\mu_k(n,s).
\end{align*}
This, in conjunction with (\ref{bound}), proves the upper bound.
\end{IEEEproof}

\begin{IEEEproof}[Proof of Theorem \ref{thm4}]
Note that
$$n\int_q^1 \frac{\mu_2(n,s)}{2(ns)^2}\, {\rm d}s=-\frac{p+\log q}{2}.$$
The claim then follows from Lemma \ref{lem3} and Proposition \ref{propbi}.
\end{IEEEproof}

\subsection{Proof of Corollary \ref{coro}}

From the definitions we have
\begin{equation}\label{eq4}
H(n, p)=\log n!-n\log n+n-D(n,p)-D(n,q),
\end{equation}
where $q\equiv 1-p$ as before. Hence, Corollary \ref{coro} is an
immediate consequence of \eqref{eq4} and Theorem \ref{thm4}.
\hfill$\blacksquare$

\section{Discussion}\label{sect4}
We have obtained asymptotically sharp and readily computable bounds
on the Poisson entropy function $H(\lambda)$.  The method also
handles the entropy of the binomial, and the relative entropy
$D(n,p)$ between the binomial$(n,p)$ and Poisson$(np)$
distributions, yielding full asymptotic expansions with explicit
constants.  While some results are of theoretical interest, bounds
on the entropy are intended to aid channel capacity calculations
for discrete-time Poisson channels, for example.

Besides entropy calculations, the method also extends to
quantities such as the fractional moments of these familiar
distributions. An example is $E[\sqrt{N_\lambda}]$, which appears as
Example~1, Entry~10, in Chapter~3 of Ramanujan's notebook
(\cite{B}). Analogous to Lemma \ref{lem2}, a double inequality can
be obtained for $E[\sqrt{N_\lambda}]$ (details omitted). This is of
some statistical interest because it gives accurate bounds on the
bias associated with the square root transformation, which is
variance-stabilizing for the Poisson.

For theoretical considerations, the Taylor formulae, integral
representations, and related results can also help to establish
interesting monotonicity properties of quantities such as
$H(\lambda)$.  For example, it can be shown that $H'(\lambda)$ is
a completely monotonic function of $\lambda$, i.e., $(-1)^{k-1}
H^{(k)}(\lambda)\geq 0$ for all $k\geq 1$.  It appears that many
entropy-like functions associated with classical distributions are
completely monotonic (\cite{Yu2}).  One conjecture
(\cite{Yu3}, Conjecture 1) states that $D(n, \lambda/n),\
n>\lambda,$ is completely monotonic in $n$ for fixed $\lambda>0$.
The method of this work may prove useful toward resolving such
conjectures.

\section*{Acknowledgments}

The authors would like to thank the Editor and the referees for
their careful reading of the manuscript and for their remarks and
suggestions, which greatly improved the final outcome.

This work has been partially supported by research grants
MTM2008-06281-C02-01/MTM, DGA E-64, UJA2009/12/07 (Universidad de
Ja\'{e}n and Caja Rural de Ja\'{e}n), and by FEDER funds.

\begin{IEEEbiographynophoto}{Jos\'{e} A. Adell} received the B.S. degree in 
mathematics from Zaragoza University, Spain, in 
1980, and the Ph.D. degree in mathematics from 
the Basque Country University, Spain, in 1983. 
During 1981--1990, he was Assistant Professor in 
the Department of Mathematics at the Basque 
Country University.  During 1991--2009, he has been 
Associate Professor in the Department of Statistics at Zaragoza University.
\end{IEEEbiographynophoto}

\begin{IEEEbiographynophoto}{Alberto Lekuona} received the B.S. degree in 
mathematics from Zaragoza University, Spain, in 
1985, and the Ph.D. degree in mathematics from 
Zaragoza University,  in 1996. During 1987--2000, 
he was Assistant Professor in the Department of 
Statistics at Zaragoza University, and Associate 
Professor at the same place in the period 2001--2009.
\end{IEEEbiographynophoto}

\begin{IEEEbiographynophoto}{Yaming Yu} (M'08) received the B.S. degree in
mathematics from Beijing University, P.R. China, in 1999, and the Ph.D.
degree in statistics from Harvard University, in 2005.  Since 2005 he has
been an Assistant Professor in the Department of Statistics at the
University of California, Irvine.
\end{IEEEbiographynophoto}

\end{document}